\documentclass[a4paper]{aa}
\usepackage{graphicx,times}

\def\martic{Marti\'c }
\def\paterno{Patern\`o }

\def\gsim{\lower.4ex\hbox{$\;\buildrel >\over{\scriptstyle\sim}\;$}}
\def\lsim{\lower.4ex\hbox{$\;\buildrel <\over{\scriptstyle\sim}\;$}}
\def\~  {$\sim$}

\def\newpage{\vfill\eject}

\def\apj{{ApJ }}
\def\apjs{{ Ap. J. Suppl. }}
\def\apjl{{ Ap. J. Letters }}

\def\aa{{A\&A }}

\def\an{{Astron. Nachr. }}
\def\sp{{Solar Phys. }}

\def\bib{\item{}}
\renewcommand{\vec}[1]{\mbox{\boldmath$#1$}}

\def\CD{Christensen-Dalsgaard }

\def\L{$\Lambda$}

\def\Oms{{\it \Omega^*}}

\def\lll{${\rm Log }\;  L/L_\odot$}
\def\llr{${\rm Log }\;  R/R_\odot$}
\def\llt{${\rm Log }\;  T_{\rm e}$}
\input epsf
\begin{document}
\title{Seismic inference of differential rotation in Procyon A}
\author{A. Bonanno \inst{1} \and M. K\"uker \inst{2} \and L. \paterno \inst{3}}
\offprints{A. Bonanno; \email abo@oact.inaf.it}
\institute{INAF -- Osservatorio Astrofisico di Catania, Via S. Sofia 78, I-95123 Catania, Italy \and
Astrophysikalisches Institut Potsdam, An der Sternwarte 16,
D-14482 Potsdam, Germany
\and
Dipartimento di Fisica e Astronomia, Sezione Astrofisica, Universit\`a di Catania, Via S. Sofia 78, I-95123 Catania, Italy}
\date{Received  / Accepted }
\abstract{The differential rotation of the  F5V-IV star Procyon A is computed for a class of models 
which are consistent with recent astrometric and asteroseismic
data.  The rotation pattern is determined by solving the Reynolds equation for motion, 
including the convective energy transport, where the latter is anisotropic owing 
to the Coriolis force action which produces a horizontal temperature gradient between the poles and the equator.   
All the models show a decrease of the rotation rate with increasing 
radius and solar-type isorotation surfaces with the equator rotating faster than the poles, the 
horizontal rotational shear being much smaller for models with a less extended convective envelope.  
The meridional flow circulation can be either clockwise or counter-clockwise, and in some cases
a double latitudinal cell appears.
The rotational splittings are  calculated  for low degree $p$-modes with $l=1,\, m=1$ and $l=2,\, m=1,2$, and it is shown that, for modes with $m=1$, the stronger is the horizontal differential rotation shear the weaker the effect on the 
average rotational splitting expected, whilst the opposite happens for the mode with $m=2$.
On the basis of the present study, a resolution of $10\,\rm nHz$ in individual oscillation frequencies seems to be necessary to test the different dynamical behaviour of the proposed models, that appears barely achievable even in the forthcoming space missions. However, the average over several observed splittings could produce the required accuracy.

} 

\maketitle
%
%
%
\section{Introduction}
%
One of the brightest stars in the sky ($m_V = 0.34$) is Procyon A ($\alpha$ CMi), hereinafter
simply called Procyon,
an extensively  studied object  whose evolutionary
phase is believed to be that of a subgiant star with a convective core, a radiative envelope 
and a thin surface convective zone, where $p$-modes are presumably excited 
(Christensen-Dalsgaard \& Frandsen 1983; Houdek et al. 1999). 

For these reasons Procyon is a favorite target for asteroseismic
observations. The first clear evidence for the presence of oscillations was obtained 
by \martic et al. (1999), but when it was observed in 2004 by the MOST space mission,  the
reported precision from the photometric measurements (15 ppm) was not sufficient for the 
detection of $p$-modes (Mattews et al. 2004).  
On the other hand, three different teams have then reported the identification of $p$-modes
on Procyon (\martic et al. 2004; Eggenberger et al. 2004; Claudi et al. 2005),
leading to the determination of the  large frequency separation in the range 
$54 \,\mu \rm Hz$ -- $56 \,\mu \rm Hz$.
It is thus conceivable that 
future observation campaigns from ground or space will further refine the determination of
the basic asteroseismic parameters and, in addition, will be able to measure
a possible rotational splitting in this    
slow rotator.  

The present work is addressed to the determination of the differential rotation for a class of models of Procyon, 
consistent with the presently available  astrometric and asteroseismic data, and the related rotational splittings.   
As the evolutionary models of Procyon show that the mass in the surface convection zone is very small,
ranging from  $ \sim 10^{-7}\,M_{\odot}$ to $ \sim 10^{-4}\,M_\odot$ during the evolution,   
it is not clear which type (if any) of differential rotation is present.  
In particular, the question we  would like to investigate is whether a significant amount of latitudinal shear is present, so that 
a solar type of differential rotation, $\Omega \sim \Omega(r,\theta)$, is generated, or instead  $\Omega \sim \Omega(r)$, as is often 
assumed in asteroseismic investigations (Di Mauro et al. 2000).  
 
Recent advances in mean-field theory of solar differential rotation have proven to be very successful 
in explaining internal solar rotation.  The solar rotation pattern can be explained in terms of angular 
momentum transport by the gas flow in the convection zone. Assuming that Reynolds stresses are the only
relevant mechanism of angular momentum transport, K\"uker et al. (1993) found a remarkably good agreement between 
the results of their model and the ``observed'' rotation, as deduced from the inversion of rotational splittings. 
Kitchatinov \& R\"udiger (1995) and K\"uker \& Stix (2001) presented a refined model that included both the 
meridional flow and the convective heat transport, and showed that the flows driven by differential rotation 
and a small horizontal temperature gradient, owing to the anisotropic convective heat transport, have opposite directions, 
and roughly balance each other in the solar convection zone.
This analysis was extended to a generic class of zero age main sequence F stars of mass $M =1.2\, M_\odot$ 
in which the convection zone was about 20\% of the stellar
radius (K\"uker \& R\"udiger 2005a) leading to the conclusion that a solar-type  rotation is the most common rotational law,
at least for moderately rotating solar type stars.  

In the case of Procyon one instead deals with a $M \sim 1.42\,M_\odot - 1.5\, M_\odot$ evolved F star  which presumably is ending or just ended 
the main sequence phase, with a convection zone much more shallow   
than that of the cases discussed in K\"uker \& R\"udiger (2005a).

%
\section{The evolutionary models}
%
The starting step of our investigation was to construct evolutionary models of
Procyon by means of the GARSOM code (Schlattl et al. 1997, 1999; Bonanno et al. 2002)  
whose structure provides the basic stratification for solving the Reynolds equation (K\"uker \& Stix 2001). 

In the present calculations the  OPAL opacities (Iglesias \&
Rogers 1996) implemented in the low-temperature regime by 
Alexander \& Fergusson (1994) tables have been used. 
As far as the equation of state is concerned, we used either the OPAL-EOS 1996 (Rogers
et al. 1996) or the MHD (Hummer \& Mihalas 1988; Mihalas et al. 1988; 
D\"appen et al. 1988) tables. The  OPAL-EOS 1996 tables have been updated by a relativistic treatment of the
electrons and by improving the activity expansion
method for repulsive interactions (Rogers \& Nayfonov 2002).
The nuclear reaction rates have been taken either from Bahcall et al. (1995) or
Adelberger et al. (1998). The convection has been treated by the classical mixing length theory. Microscopic diffusion of hydrogen, helium and all the major metals has
been taken into account. The outer boundary conditions have been determined by assuming an Eddington gray atmosphere.  

The most recent determinations of 
the astrometric parameters of the visual binary orbit of Procyon 
have been obtained by means of the Multichannel Astrometric Photometer observations
during the period 1986--2004. The inferred mass of the primary component is  
found to be $M=1.43 \pm 0.034\,M_\odot$ (Gatewood \& Han 2006),  
whilst a previous determination by 
Girard at al. (2000) reported  $M=1.497 \pm 0.037\, M_\odot$, with a 
parallax $\Pi=283.2\pm 1.5\, \rm mas$. Although the two measurements are consistent 
within the error bars, 
if only the Wide Field and Planetary Camera 2 observations are used in combination with the 
Hipparcos parallax, $ \Pi= 285.93 \pm 0.88 \rm\,mas$, the resulting mass is 
$M=1.42 \pm 0.04\, M_\odot$, much closer to the Gatewood \& Han (2006) value.  
Here we shall assume the conservative range  $1.42\, M_\odot$ - $1.50\, M_\odot$
to constrain the mass of Procyon in our models, although the average value of these measurements
would give a mass $M=1.45 \pm 0.02\, M_\odot$.    

The luminosity has been estimated from the mean visual magnitude obtained  by Allende Prieto et al. (2001) and 
bolometric correction by Flower (1996), so that 
$\log\,L/L_\odot = 0.84 \pm 0.02$. Recent measurements of the angular diameter by Kervella et al. (2004)    
lead to $ R/R_\odot = 2.048 \pm 0.025$, while the parallax of Girard et al. (2000) gives 
$ R/R_\odot = 2.067 \pm 0.028$.  The effective temperature has been calculated by Fuhrmann et al. (1997) who found
$\log T_{\rm e} = 3.815 \pm 0.006$. 

The metallicity of Procyon, as derived from spectroscopic measurements, results to be close to the solar one
(Takeda et al. 1996; Kato et al. 1996), and for simplicity we adopted the same
(standard) metallicity mixture of the Sun (Grevesse et al. 1996) so that at the surface $(Z/X)_{\rm ph}=0.0245$. 
The inclusion of metal diffusion in evolutionary models of Procyon leads to an almost complete depletion of 
the metal content at the surface (Chaboyer et al. 1994).
The reason is that the gravitational settling is expected to take place below
the convection zone which is only $ \sim 10^{-4}\, M_\odot$ thick.
In order to avoid this problem, we have considered a mass loss according with the semi-empirical Reimers (1975)
prescription, $\dot{M} = -4\times 10^{-13}\, \eta \, (L/gR)$, where the surface gravity, $g$, luminosity, $L$, and radius, $R$, are expressed in solar units, and $\eta$ 
is an efficiency parameter, whose value depends on the evolutionary phase of the star from the 
main sequence to the red giant branch; here a value close to 0.1 results to be adequate.  
The rotational velocity, as estimated by Allende Prieto et al. (2002), is    
$v \sin \, i = 2.7 \pm 1.0 \, \rm km \, s^{-1}$. If we assume that the rotational 
axis of Procyon is perpendicular to the plane of visual orbit 
($i= 31.1^{\circ} \pm 0.6^{\circ}$, Girard et al. 2000), the surface rotational velocity
should be $5.2 \pm 1.9 \, \rm km \, s^{-1}$, which corresponds to a rotational period of about $20\,\rm d$.
On other the hand, in the extreme case of $i= 90^{\circ}$ the rotational period is about $33\,\rm d$. 
Present asteroseismic data are not accurate enough to put strong constraints on the 
computed models, although  several groups have succeeded in measuring the 
large frequency separation $\Delta \nu_0$. \martic et al. (2004) found $\Delta \nu_0 = 53.6 \pm 0.5 \, \mu \rm Hz$,   
while Eggenberger et al. (2004) $\Delta \nu_0 = 55.5 \pm 0.5 \, \mu \rm Hz$. More   
recent measurements, obtained with the high resolution spectrograph SARG operating with the $3.5\,\rm m$ Italian Galileo Telescope at Canary Islands,
report instead 
$\Delta \nu_0 = 56.2 \pm 0.5 \, \mu \rm Hz$ (Claudi et al. 2005). 
We thus constructed evolutionary models
of Procyon constrained by $54\,\mu\rm Hz \le \Delta \nu_0 \le 56\,\mu\rm Hz$, 
and computed four representative models with masses in the range 
$1.42\,M_\odot$ - $1.5\,M_\odot$.
In all the models, the  
the mixing length parameter, initial helium and metal
abundances, diffusion of the elements, and the mass loss parameter $\eta$ have been adjusted in order to reproduce the observed luminosity  
and effective temperature within a box delimiting their determination uncertainties, with the above mentioned constraint on $\Delta \nu_0$. As far as the variation of the mass loss parameter $\eta$ is concerned, an exploration in the range $0.05\leq\eta\leq 0.15$, that is still consistent with the asteroseismic requirements of matching the value of $\Delta \nu_0$, does not produce sensible effects on the stratification of the convection zone and by consequence on its dynamical behaviour.

\begin{figure}
\includegraphics[width=9.0cm]{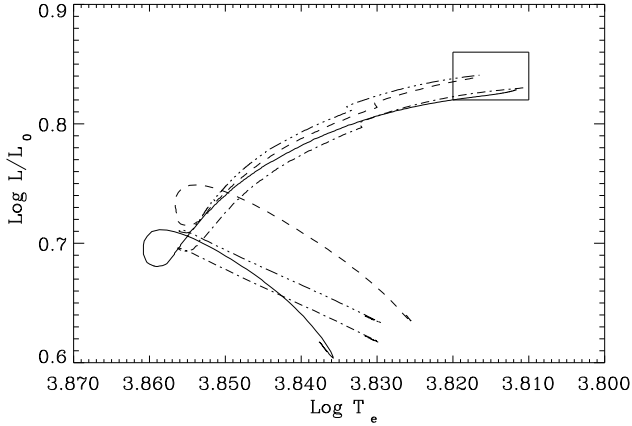}
\caption{The evolutionary tracks of the models summarized in Table \ref{tab1}. The solid line is for Model 1, the dashed for Model 2,
the dot-dashed for Model 3, and the dot-dot-dashed for Model 4. The box indicates the uncertainties in the
observation determinations of luminosity and effective temperature.
\label{track} }
\end{figure}

The basic characteristics of the four representative models are summarized in Table \ref{tab1}, while the evolutionary tracks of the models are presented
in Fig. \ref{track}. 
\begin{table*}
\caption{Characteristic quantities of the four Procyon models. The
suffixes ph and cz indicate top
(photosphere) and bottom of 
convective envelope, respectively. 
Model 1 includes diffusion of H and He only, while
Models 2, 3 and 4 include diffusion of all the elements.}
\begin{flushleft}
\begin{tabular}{*{13}{c}}
\hline
\hline
Model & $M/M_\odot$ & Age [Gry] & \lll & \llr & \llt & $Z_{\rm ph}$ & $Y_{\rm ph}$ & ${R_\mathrm{cz}}/{R_\mathrm{s}}$ & 
$\Delta \nu_0 \; [\mu {\rm Hz}] $ & $\eta$ \\
\hline  
1& 1.42&  2.2 & 0.829& 0.315& 3.812& 0.01805&  0.237& 0.856&   54.1 & 0.1\\
2& 1.45&  2.1 & 0.830& 0.315& 3.810& 0.01828&  0.241& 0.867&  54.5 & 0.1\\
3& 1.48&  1.8 & 0.839& 0.310& 3.816& 0.01845&  0.215& 0.918&  55.4 & 0.1\\
4& 1.50 & 1.9 & 0.847& 0.310 & 3.816 & 0.01798& 0.228 & 0.919 & 55.0  & 0.1\\
\hline
\end{tabular}
\end{flushleft}
\label{tab1}
\end{table*}
Note that the
Model 2 represents the model with a mass corresponding to the average mass of the most recent determination, as already discussed. 

%
\section{Differential rotation and heat transport in the outer convection zone}
%
The solar differential rotation was successfully reproduced by means of a mean-field model based on the Reynolds stress transport of angular momentum  (K\"uker et al. 1993), improved by the inclusion of meridional flow and turbulent heat transport  (Kitchatinov \& R\"udiger 1995). The inclusion of the above mentioned ingredients turned out to be crucial to obtain results in agreement with the isorotation surfaces deduced from helioseismology, as the Taylor-Proudman theorem predicts cylindrical surfaces of isorotation in an isothermal fluid at high Taylor numbers. Since the Taylor number assumes values as large as $10^8$ in the lower part of the solar convection zone, owing to its inviscid character, the  state of rotation in cylinders would have been unavoidable (R\"udiger et al. 1998). The model has been applied to several types of stars, either fully convective or with outer convection zones (Kitchatinov \& R\"udiger 1999; K\"uker \& Stix 2001; K\"uker \& R\"udiger 2005a,b).

\begin{figure}
\begin{center}
\includegraphics[width=8.0cm,height=6cm]{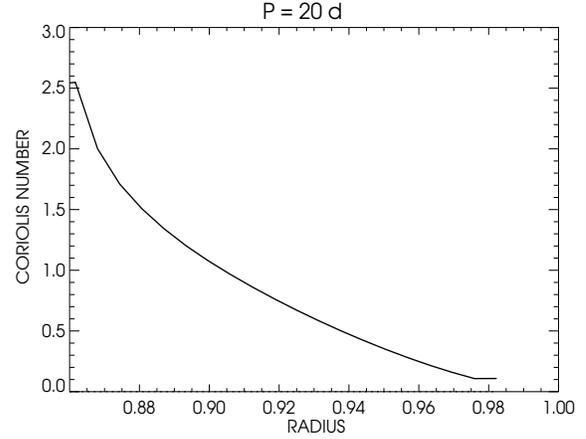}
\caption{Coriolis number as a function of the stellar fractional radius for a 
star with $M= 1.45\,M_\odot$ in the subgiant phase for a rotation period of $20\,\rm d$.}
\label{cor}
\end{center}
\end{figure}

A relevant quantity to understand the role of the turbulent transport of the angular momentum
is the Coriolis number $\Oms=2\tau \Omega$, where $\Omega$ is the stellar angular velocity and $\tau$ the characteristic turnover time of convective eddies (R\"udiger 1989).
The radial dependence of  
$\Oms$ in the outer layers of the star for a  representative model of Procyon 
is shown in Fig. \ref{cor}.
Since the inhomogeneous \L-effect depends on $\Oms^{-1}$ while the 
anisotropic \L-effect depends on $\Oms^{-3}$, it is then clear that they are both essential  
to correctly describe the turbulent transport of angular momentum (Kitchatinov \& R\"udiger 
1995, 1999).

%
\subsection{Reference model}
%
Since the influence of rotation and large-scale meridional flow on the structure of the star is very small we treat them 
as perturbations of an unperturbed, non-rotating, spherically symmetric reference state, that is a stellar convection zone in hydrostatic equilibrium, 
as computed by the stellar evolution code above described.
In order to compute the differential rotation, we simplify the reference state by assuming a perfect gas with constant specific entropy and total heat flux throughout 
the convection zone. The differential equations for temperature and gravity (Poisson equation for a spherically symmetric gravitational potential with $g(r) = -{\rm d}\Phi/{\rm d}r$) are those described in Kitchatinov \& R\"udiger (1995) for an analogous problem:
\begin{equation} \label{simptemp}
  \frac{{\rm d}T}{{\rm d}r}=-\frac{g(r)}{c_p} 
\end{equation}
\begin{equation} \label{simpgrav}
  \frac{{\rm d}g(r)}{{\rm d}r}=-2\frac{g(r)}{r}+4 \pi G \rho
\end{equation}
with the adiabatic relationship:
\begin{equation} \label{simpdens}
  \rho=\rho_e \left(\frac{T}{T_e}\right)^{1/(\gamma-1)} \\
\end{equation} 
where $c_p = {\cal R}/\nabla_{\rm ad}$ is the specific heat at constant pressure, with $\cal R$ the gas constant and $\nabla_{\rm ad}$ the adiabatic gradient, $g(r)$ the gravity, 
$\gamma = (1-\nabla_{\rm ad})^{-1}$ the adiabatic exponent, and the suffix $e$ indicates the values of the quantities at the outer boundary.
The solution is determined by choosing a reference point from the model produced by the stellar evolution code, which provides the initial values necessary for the integration.
Equations are then integrated outwards and inwards from the starting point at which the values from the reference model
have been chosen, the inward integration being terminated at the point where the radiative flux reaches the total heat flux, which coincides with
the bottom of the convection zone, and the outward integration being terminated at the stellar radius as determined by the reference model. Thus Eqs. (\ref{simptemp}) and (\ref{simpgrav}) constitute an initial value problem. To adjust the depth of the convection zone to the value of the reference model, we take the adiabatic gradient as input parameter. 
For fully ionized monatomic gas, $\nabla_{\rm ad} = 0.4$. In the outer parts of the convection zones, however, it is smaller owing to the partial ionization. The stratifications found by the stellar evolution code for the models of Procyon we considered show values of $\nabla_{\rm ad}$ as low as 0.1 in the outer parts of the convection zone. Since we assumed it constant with radius in our simple model, the values of $\nabla_{\rm ad}$ that give the best fit to the evolutionary models lie around 0.35.
We thus produce a convection zone with adiabatic stratification that closely resembles the one reproduced by the evolutionary models, that is used as the unperturbed, reference state for the computation of the differential rotation.
In spite of the simplifications used, the deviations of our adiabatic stratification from that of the evolutionary model are within a few percent in the relevant quantities everywhere in the convection zone.

%
\subsection{Heat transport}
%
Following the procedure described in R\"udiger \& Hollerbach (2004), the mean-field equation of convective heat transport is given by the following expression:
\begin{equation} \label{heat}
\rho T \frac{\partial \bar{s}}{\partial t}
          = - \nabla \cdot  (\vec{F}^{\rm conv} + \vec{F}^{\rm rad}) +\rho c_p \vec{\bar{u}} \cdot \vec{\beta},
\end{equation}
where $\bar{s}$ is the mean specific entropy. The transport terms on the RHS of Eq. (\ref{heat})
 are the convective heat flux, $F_{i}^{\rm conv} = \rho c_p \langle {u_i'} T'\rangle = \rho c_p \chi_{ij}\beta_j$, the radiative heat flux, $F_i^{\rm rad} = - 16 \sigma T^3\nabla_i T/3 \kappa \rho $, and the heat advection by meridional flow, with $\bar{\vec u}$ the mean velocity field, where $\kappa$ is the opacity, $\sigma$ the Stefan-Boltzmann constant, and $\vec{\beta}= \vec{g}/{c_p}- \vec{\nabla}{T} = - \nabla \delta {T}$ the super-adiabatic gradient. As the reference state has constant entropy $s_0$, we define the perturbation as $\delta {s} = \bar{s} - s_0$, and, since all derivatives of $s_0$ vanish, we can rewrite Eq. (\ref{heat}) in terms of $\delta {s}$:
\begin{equation} \label{heat2}
\rho T \frac{\partial \delta {s}}{\partial t}
          = - \nabla \cdot  (\vec{F}^{\rm conv} + \vec{F}^{\rm rad}) +\rho c_p \vec{\bar{u}} \cdot \vec{\beta}.
\end{equation}
On applying the mass conservation law, $\nabla \cdot (\rho \vec{\bar{u}}) = 0$,
replacing the temperature perturbation with the entropy perturbation,
$\delta s = c_p \delta T/T$, we can rewrite the last term in Eq. (\ref{heat2}) as
$\rho c_p \vec{\bar{u}} \cdot \vec{\beta} = - \nabla \cdot (\rho T \vec{\bar{u}} \delta {s})$.
Since we find in some cases that the heat transport by the meridional flow exceeds that of the convective flux, we introduce a horizontally-averaged entropy perturbation:
\begin{displaymath} 
  \delta {\bar{s}}_1 = \frac{1}{2}\int_0^\pi \delta {s} \sin \theta { d}\theta
\end{displaymath}
thus  splitting $\delta {s}$ into the two terms $\delta {\bar{s}}_1(r)$ and $\delta s_2(r,\theta)$.
We then drop out $\delta \bar{s}_1$ from the advection term to prevent convective instabilities in our system, and obtain the equation of heat transport in its final form:
\begin{equation} \label{deltas}
  \rho T \frac{\partial \delta {s}}{\partial t}
          = - \nabla \cdot  (\vec{F}^{\rm conv} + \vec{F}^{\rm rad} + \rho T \vec{\bar{u}} \delta {s_2})
\end{equation} 

%
\subsection{Equation of motion}
%
The equation of motion for the mean velocity field $\vec{\bar{u}}$ is (cf. R\"udiger 1989):
\begin{equation} \label{reynolds}
  \rho \left [ \frac{\partial \vec{\bar{u}}}{\partial t}
      + (\vec{\bar{u}}\cdot \nabla)
       \vec{\bar{u}} \right ] = -  \nabla \cdot \rho Q
           - \nabla P + \rho \vec{g}
\end{equation}
where $Q$ is the one-point correlation tensor of fluctuating velocities
$Q_{ij} =  \langle u_i' u_j' \rangle$.
On assuming axial symmetry, the azimuthal component of Eq. (\ref{reynolds}) can be written as:
\begin{equation} \label{omega}
    \rho r^2 \sin^2 \theta \frac{\partial \Omega}{\partial t}
     + \nabla \cdot \vec{t} = 0
\end{equation}
where $\vec{t} =  r \sin \theta (\rho r \sin \theta \Omega \vec{\bar{u}}^{\rm m}
    + \rho \langle u_{\phi}' \vec{u}' \rangle )$.
The two remaining components of Eq. (\ref{reynolds}) describe the meridional circulation $\vec{\bar{u}}^{\rm m}$, and can be represented by a single scalar function, the stream function $A$:
\begin{equation}
  \vec{\bar{u}}^{\rm m} = \left ( \frac{1}{\rho r^2 \sin \theta}
                           \frac{\partial A}{\partial \theta},\;
                           -\frac{1}{\rho r \sin \theta}
                           \frac{\partial A}{\partial r},\;  0 \right)
\end{equation}
On taking the azimuthal component of the {\em curl} of Eq. (\ref{reynolds}) we obtain:
\begin{eqnarray}
 \label{curl} 
 \frac{\partial \omega}{\partial t}= &-& \left[ \nabla \times 
           \frac{1}{\rho}\nabla(\rho Q) \right ]_\phi
           +r \sin \theta \frac{\partial \Omega^2}{\partial z} \nonumber \\
          &+& \frac{1}{\rho^2}(\nabla \rho \times \nabla p)_\phi + \dots 
\end{eqnarray}
where $\omega = ( \nabla \times \vec{\bar{u}})_\phi$ is the azimuthal component of the mean vorticity,  ${\partial}/{\partial z}=\cos \theta \cdot
            {\partial}/{\partial r}-{\sin \theta}/{r} \cdot
            {\partial}/{\partial \theta}$ represents the gradient along the axis
of rotation, and dots indicate second-order terms of minor importance.
As the reference state is in hydrostatic equilibrium, ${\vec {\nabla}} P = - \vec{g} \rho$,
we can rewrite the baroclinic term on the RHS of Eq. (\ref{curl}) in terms of specific entropy:
\begin{equation}
\label{baroclinic}
  \frac{1}{\rho^2}(\nabla \rho \times \nabla P)_\phi =
 - \frac{1}{c_p \rho}(\nabla s \times \nabla P)_\phi \approx
 - \frac{g}{r c_p} \frac{\partial s}{\partial \theta}
 \end{equation}
On neglecting the second-order terms, we have three terms on the RHS of Eq. (\ref{curl}); 
the first, which contains the Reynolds stresses, is purely viscous, while the second and third 
drive the meridional flow. Close to the poles, the gradient of the rotation rate along the rotation axis dominates, as the horizontal entropy gradient vanishes there. At low latitudes, on the other hand, ${\partial \Omega}/{\partial z}$ vanishes and the baroclinic term dominates.  As both terms can be either positive or negative, they can in principle enhance or cancel each other. For solar-type stars it has been found that both terms balance each other in a wide range of rotation rates, leading to a slow meridional flow and a strong differential rotation (K\"uker \& R\"udiger 2005).

%
\subsection{Transport coefficients}
%
Instead of the standard mixing-length theory we use the expressions derived by Kitchatinov \& R\"udiger (1999) for the convective heat transport and Reynolds stresses. The heat conductivity tensor, $\chi_{ij}$, can be written as the product of a scalar diffusion coefficient, $\chi_t$, and a dimensionless tensor, $\Phi_{ij}$, so that $\chi_{\rm ij} = \chi_t \Phi_{ij}$,
where $\chi_{\rm t} =  c_\chi{\tau_{\rm c} g \alpha^2 H_p^2}\langle \beta_r \rangle/{4\,T} $, with $\tau_c$ the turbulent eddy correlation time, $\alpha$ the classical mixing-length parameter, $H_p$ the pressure scale height, $c_\chi$ a dimensionless coefficient for heat transport, and the triangular brackets averages over $\theta$.
The stress tensor has the general form ${\cal T}_{ij}= - \rho Q_{ij}$, with the correlation tensor:
\begin{equation}
 Q_{ij}= -{\cal N}_{ijkl}\frac{\partial \Omega_k}{\partial x_l} + \Lambda_{ijk} \Omega_k.
\end{equation}
The first term on the RHS vanishes for rigid rotation and represents an anisotropic viscosity; the second one, the \L-effect, is non-zero for rigid rotation, and therefore causes differential rotation unless balanced by the meridional flow.
We write the viscosity tensor, ${\cal N}_{ijkl} = \nu_t \Psi_{ijkl}$, in a similar way as the convective heat transport, namely $\nu_{\rm t} = c_\nu {\tau_{\rm c} g \alpha^2 H_p^2}\langle \beta_r \rangle/{4\,T}$, where $c_\nu$ is a dimensionless coefficient for viscous transport.
The \L-effect appears only in two components of the correlation tensor, and it can be written as follows:
\begin{eqnarray}
  Q^\Lambda_{r \phi} &=& \nu_t V \sin \theta \Omega, \hspace{5mm} V=V^0+V^1 \sin^2 \theta \\
  Q^\Lambda_{\theta \phi} &=& \nu_t H \cos \theta \Omega, \hspace{5mm} H=V^1 \sin^2 \theta.
\end{eqnarray} 
The tensors $\Phi_{ij}$ and $\Psi_{ijkl}$ contain the anisotropy of the convective heat transport and the viscous part of the stress tensor, as computed by Kitchatinov et al. (1994). Together with the \L-effect coefficients $V^0$ and $V^1$ (Kitchatinov \& R\"udiger 1993, 2005)  determine the transport of heat and angular momentum by convection. Note that the updated \L-effect coefficients of Kitchatinov \& R\"udiger (2005) remove the discrepancy between theoretical and observed angular velocity distribution in the outer $30,000\,\rm km$ of the solar convection zone found in models which uses the older expressions. 
To adjust the model for the updated expressions for the \L-effect, we now use $c_\nu = 0.27$ with $\alpha = 2$ rather than the values from K\"uker \& R\"udiger (2005a,b), who used $c_\nu = 0.15$ with $\alpha=5/3$.  With a Prandtl number, $\rm P_r = \nu_t/\chi_t = 0.8$ (Kitchainov et al. 1994), there is a very good agreement between the computed and observed rotation pattern for the Sun. The transport coefficients have been calibrated with the Sun, whose differential rotation is known from helioseismology. Numerical simulations on the effect of changes of transport coefficients show that the strength of horizontal rotational shear is quite insensitive to changes of parameters, while meridional flow is more sensitive leading to slight changes in the rotational pattern. 
Both the existence of the \L-effect and the anisotropy of the convective heat transport have been confirmed by simulations (K\"apyl\"a et al. 2004; R\"udiger et al. 2005). 

%
\subsection{Boundary conditions and gravity darkening}
%
The numerical code we used to solve Eqs. (\ref{deltas}), (\ref{omega}), and (\ref{curl}) is the same as that used by K\"uker \& Stix (2001). It uses a second-order finite difference scheme that conserves angular momentum and energy. The equations for the reference model have been solved by using a fourth-order Runge-Kutta scheme. 
In the equation of motion, we impose stress-free boundary conditions on both boundaries. In the heat transport equation, we fix the total luminosity on both boundaries, and assume that is independent of latitude, so that $L = 4\pi r^2 F$.
This fixes the radial component of the entropy gradient, but not the entropy itself, so not precluding a horizontal entropy gradient. As we have written the heat transport equation as a conservation law, these boundary conditions ensure that a stationary state is eventually reached.

Our code also includes the option of taking into account the gravity darkening by adding a rotation-dependent term to the heat flux equation:
\begin{equation}
 F =  \frac{L}{4 \pi r^2} \left[1 + \frac{2 \varepsilon}{3+\varepsilon} \left(\frac{3}{2}\cos^2\theta-\frac{1}{2}\right)\right]
\end{equation}
where $\varepsilon=\Omega^2r^3/GM$ (R\"udiger \& K\"uker 2002).
For a star like Procyon with a rotation period of $20\,\rm d$, a value of $\varepsilon = 1.8 \times 10^{-4}$ is found, small value that cannot affect luminosity. 
In fact we did not find a significant difference between runs with and without gravity darkening.

%
\section{Rotational splittings}
%
The rotational splittings of oscillation frequencies depend on the radial and latitudinal
angular velocity distribution in the interior of the star.
It could be of interest, essentially in the light of possible future observations,
to calculate the rotational splittings for low-degree $p$-modes
expected for our differential rotation models of Procyon.
We use the following parameterization for the rotation law:
\begin{equation}
\Omega(r,\theta) = \sum_{i=0}^{i_{\rm max}}\Omega_i(r)\cos^{2i}\theta
\end{equation}
so that the angular integration can be explicitly performed, and the 
rotational splittings are (Cuypers 1980): 
\begin{equation}
\delta\nu_{nlm} = \frac{m}{2\pi}\sum_{i=0}^{i_{\rm max}}
\int_0^R K_{nlmi}(r)\Omega_i(r)dr 
\end{equation}
with the kernels $K_{nlmi}$ (\CD 1997):
\begin{eqnarray}
&&K_{nlmi}=\nonumber\\
&&\rho r^2 \Big \{ \Big \{ (\xi_r-\xi_h)^2+
(l(l+1)-2i^2-3i-2)\xi_h^2\Big ] Q_{lmi}\nonumber\\
&&+i(2i-1)Q_{lmi-1} \Big\}/{\cal I}
\end{eqnarray}
where:
\begin{equation}
{\cal I} = \int_0^R \rho r^2 (\xi_r^2+l(l+1)\xi_h^2) dr 
\end{equation}
and,
\begin{equation}
Q_{lmi}=\frac{2l+1}{2}\frac{(l-|m|)!}{(l+|m|)!}\int_{-1}^{1}x^{2i}[P_l^m(x)]^2dx
\end{equation}
$\xi_r$ and $\xi_h$ being the radial and horizontal components of the eigenfunctions, respectively. In our investigation the functions $\Omega_i(r)$ have been determined from the numerical integration 
of the Reynolds equation (\ref{reynolds}) with $i_{\rm max}=4$. 

%
\section{Results}
%
%
\begin{figure} \mbox{
 \includegraphics[width=4.4cm]{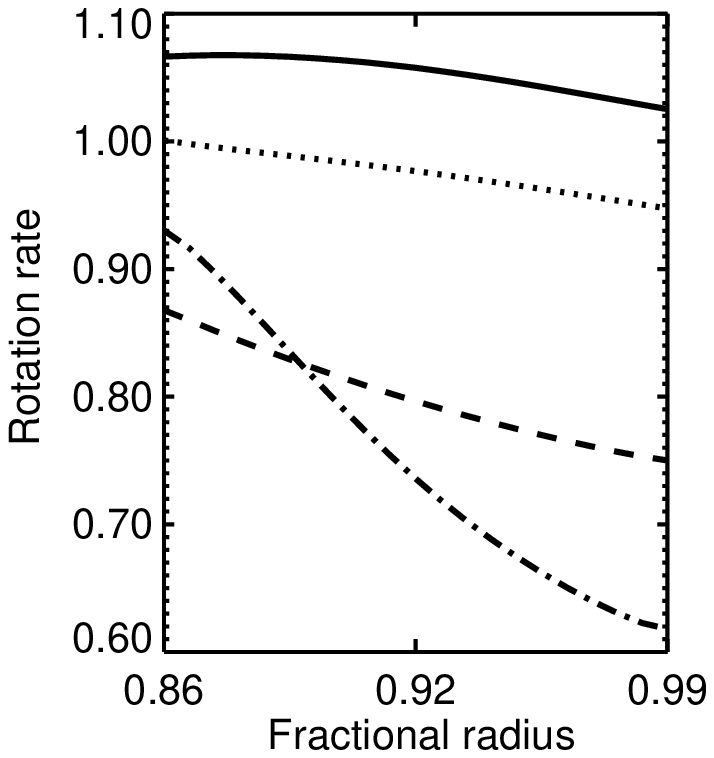}
 \includegraphics[width=4.4cm]{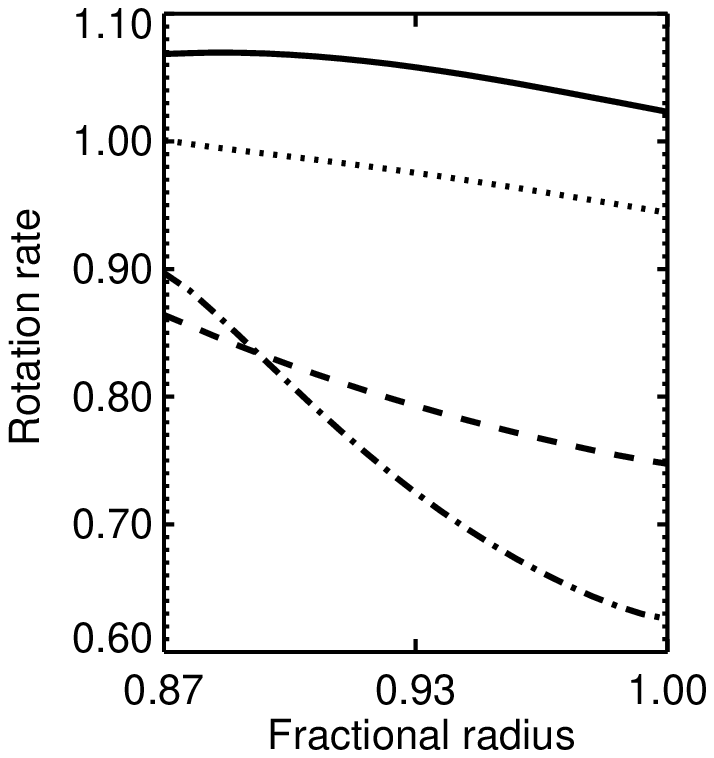} 
} 
 \\ \mbox{
 \includegraphics[width=4.4cm]{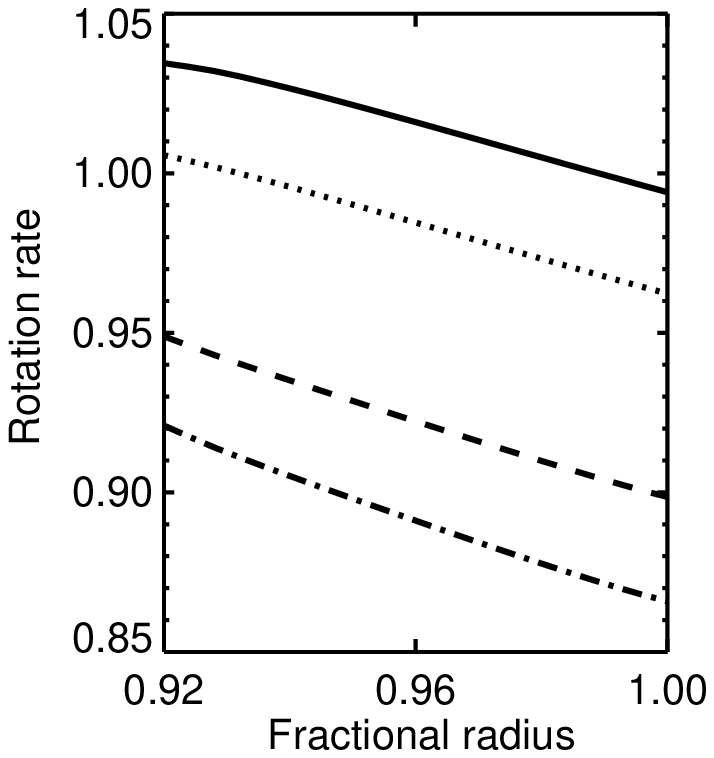}
 \includegraphics[width=4.4cm]{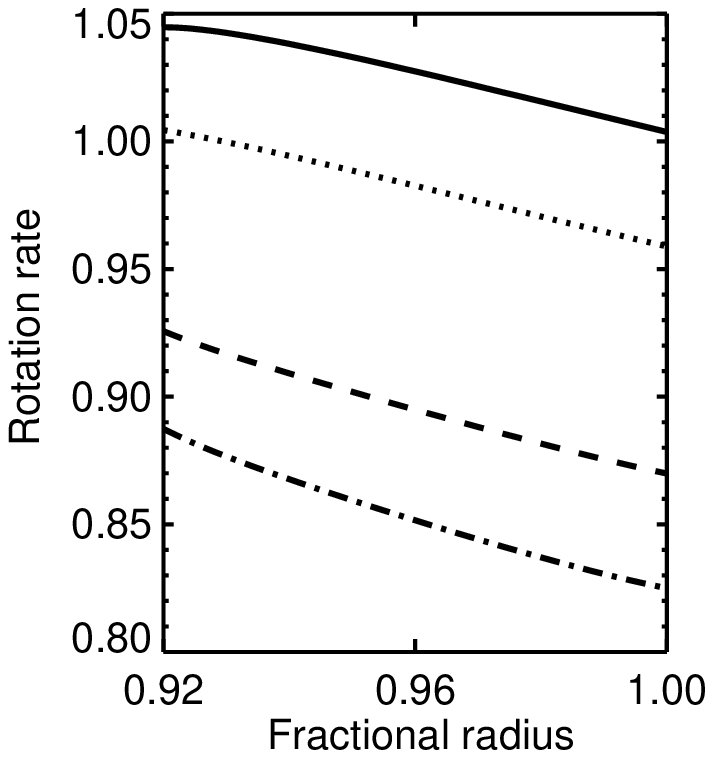} 
}
\caption{ \label{p20}
The normalized angular velocity profiles for the four models with a rotation period $P_{\rm rot}= 20\,\rm d$. The four panels show respectively: top left: Model 1; 
top right: Model 2; bottom left: Model 3; bottom right: Model 4. 
The different line-styles denote different latitudes: solid: equator; dotted: $30^\circ$; 
dashed: $60^\circ$; dash-dotted: poles. The normalization of the angular velocity is done with respect to an average angular velocity $\bar{\Omega}$ as described in the text.}
\end{figure}
%
\begin{figure} 
 \mbox{
 \includegraphics[width=3.0cm]{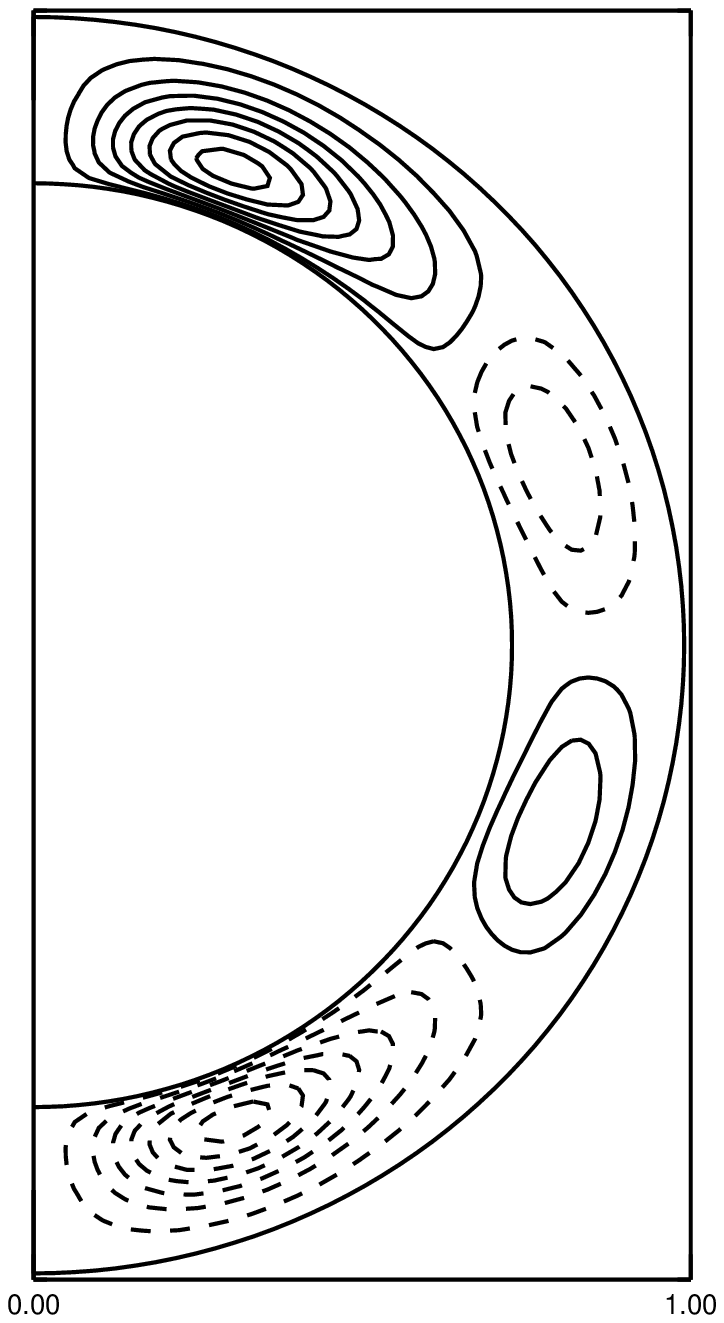}
 \includegraphics[width=3.0cm]{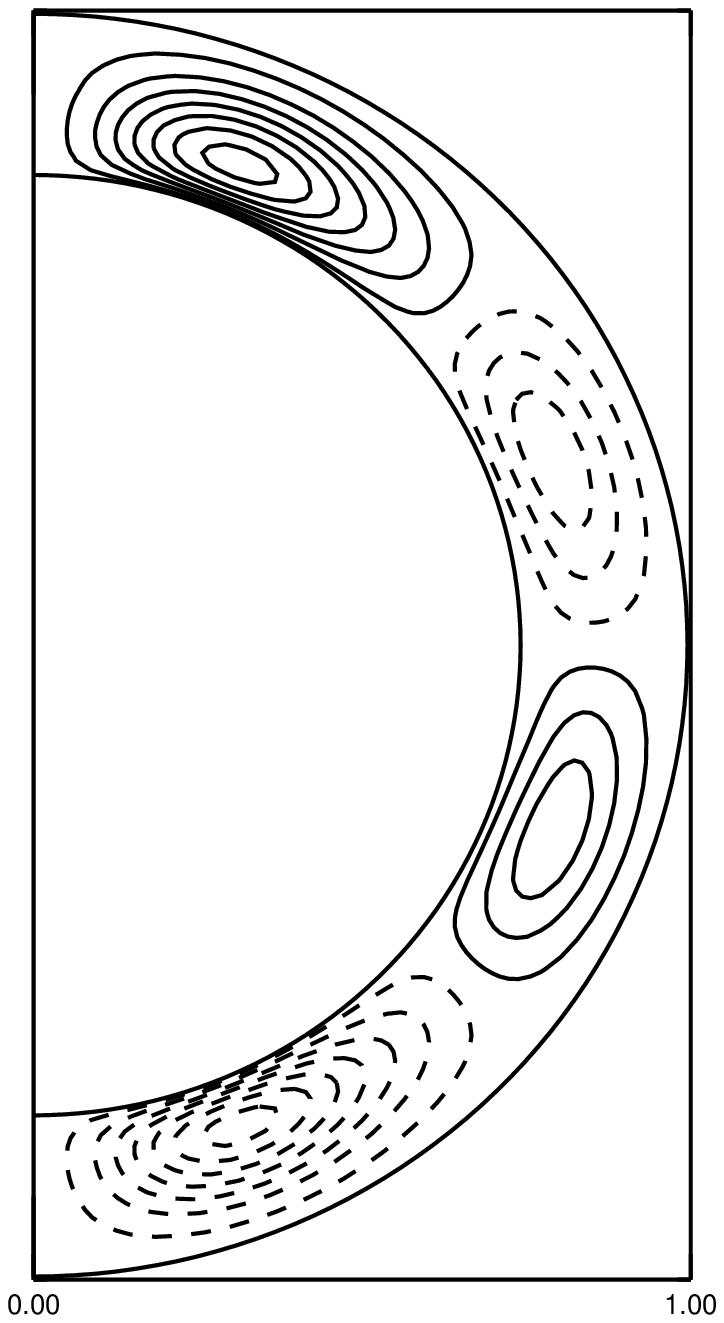} 
} 
   \\  \mbox{
 \includegraphics[width=3.0cm]{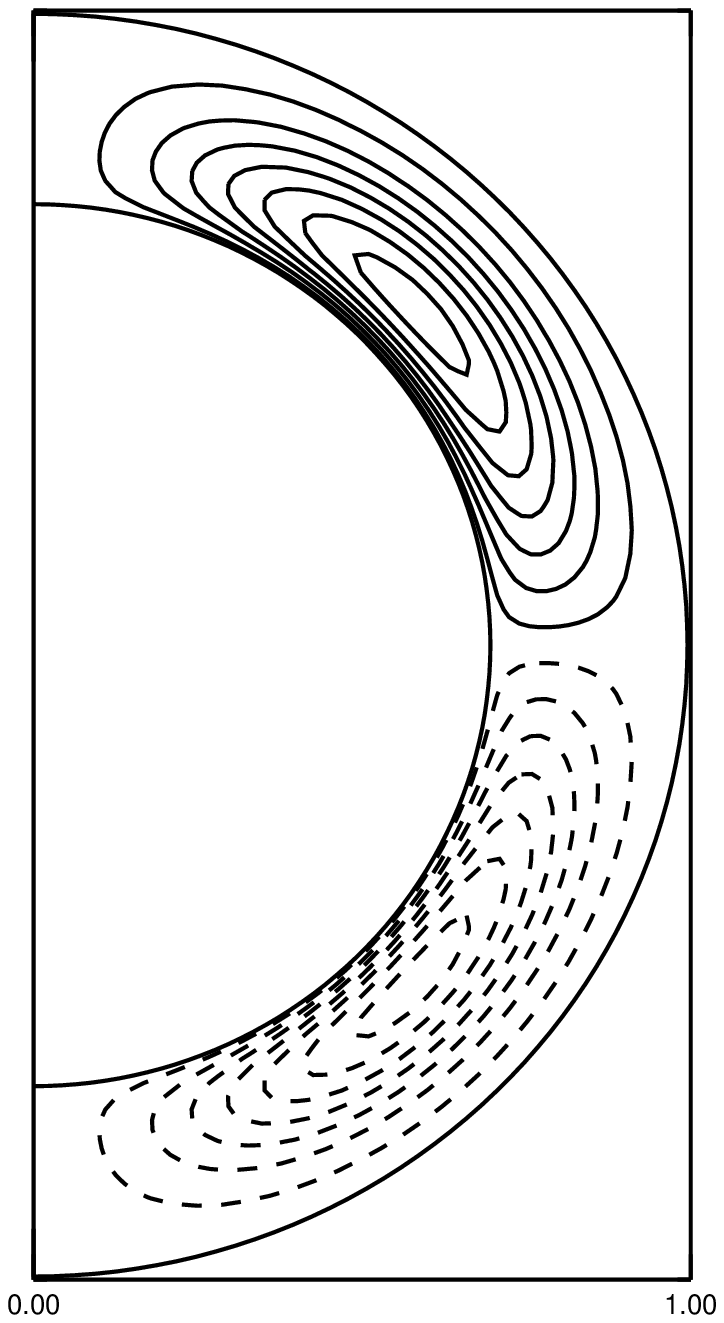}
 \includegraphics[width=3.0cm]{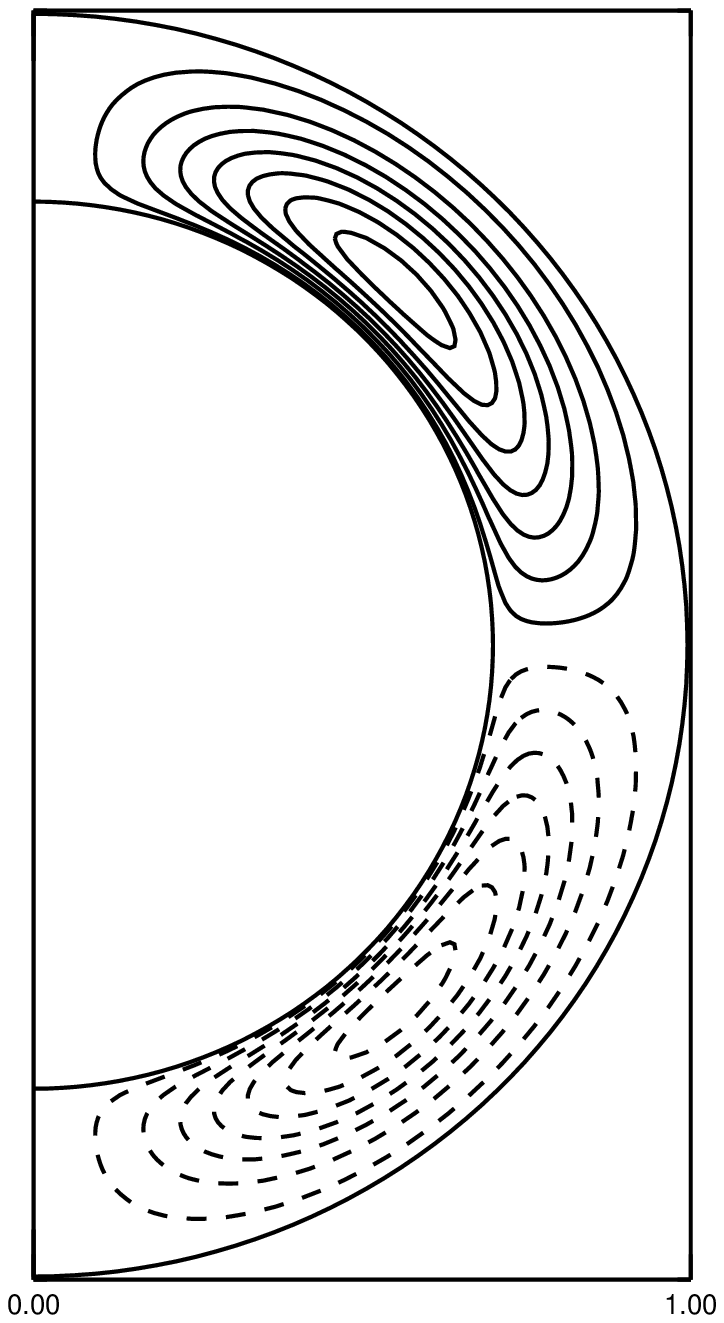} 
}
\caption{ \label{p20stream} The meridional flow patterns corresponding to the rotation profiles in Fig. \ref{p20}. The four panels show respectively: top left: Model 1; 
top right: Model 2; bottom left: Model 3; bottom right: Model 4. 
The solid lines of iso-velocity denote clockwise and the dashed ones counter-clockwise motions. 
For clarity, in the diagrams 
the depth of the convection zone has been enlarged by a factor of 2 for Models 1 and 2 and 4 for Models 3 and 4.
The meridional motion maximum speeds at the stellar surface are respectively: $100\,\rm m\,\rm s^{-1}$ for Model 1; $87\,\rm m\,\rm s^{-1}$ for Model 2; $41\,\rm m\,\rm s^{-1}$ for Model 3; $37\,\rm m\,\rm s^{-1}$ for Model 4.}
\end{figure}
%
\begin{figure}
 \mbox{
 \includegraphics[width=4.4cm]{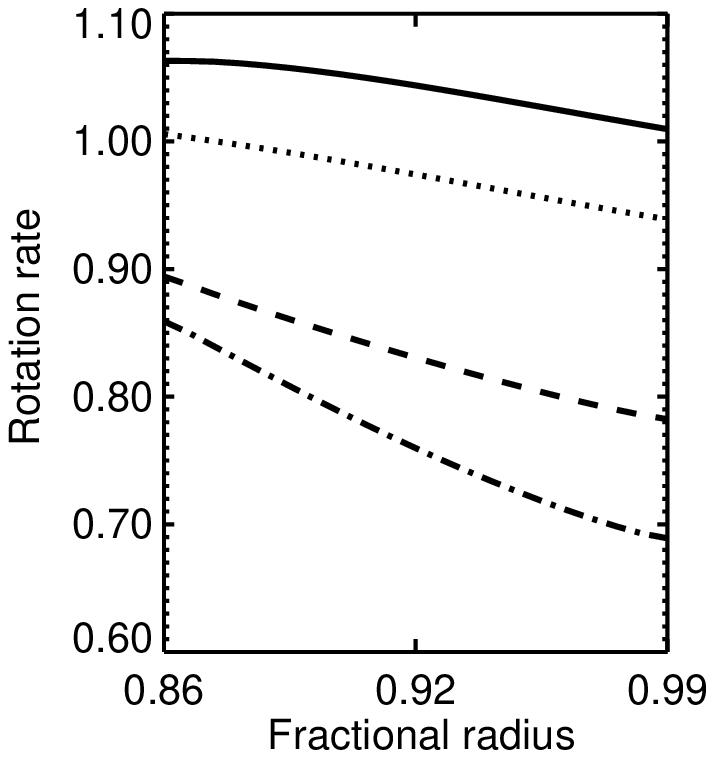}
 \includegraphics[width=4.4cm]{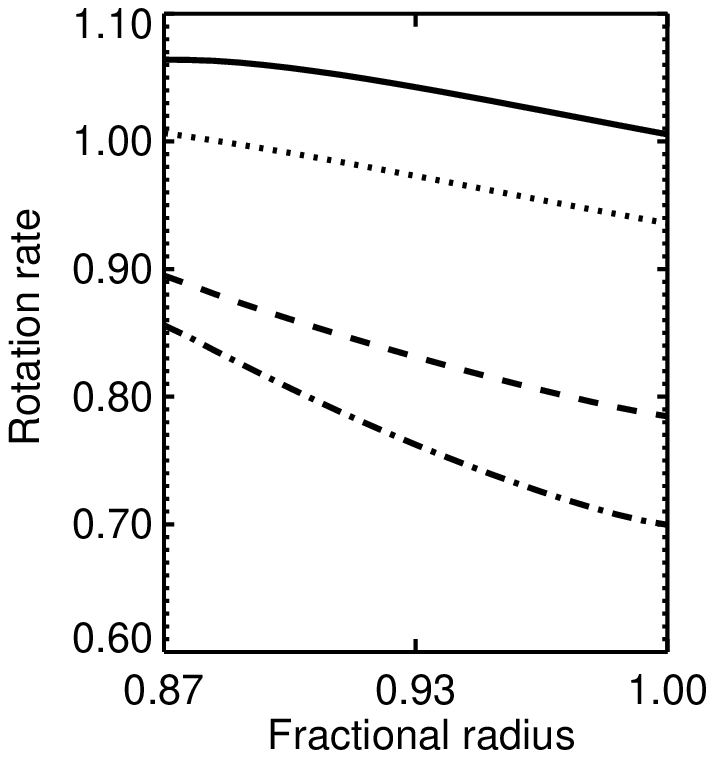} 
} 
 \\ \mbox{
 \includegraphics[width=4.4cm]{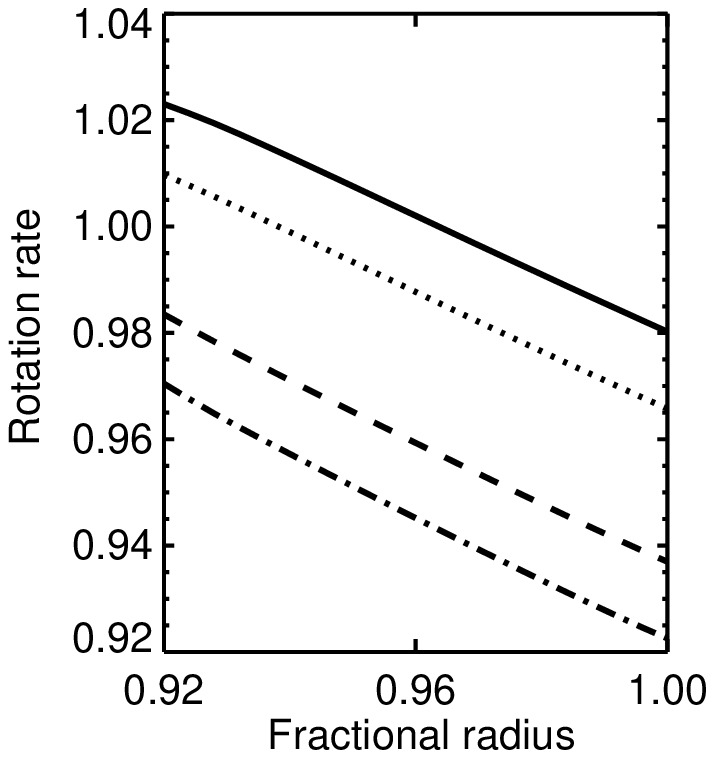}
 \includegraphics[width=4.4cm]{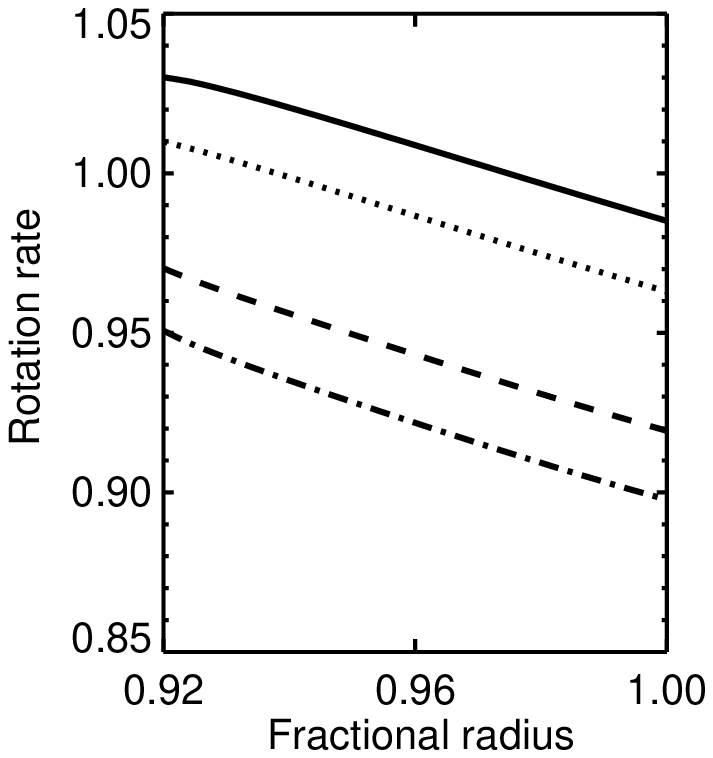} 
}
\caption{ \label{p33} The same as in Fig. \ref{p20} but for a rotation period $P_{\rm rot} = 33\,\rm d$.}
\end{figure}
%
\begin{figure}
 \mbox{
 \includegraphics[width=3.0cm]{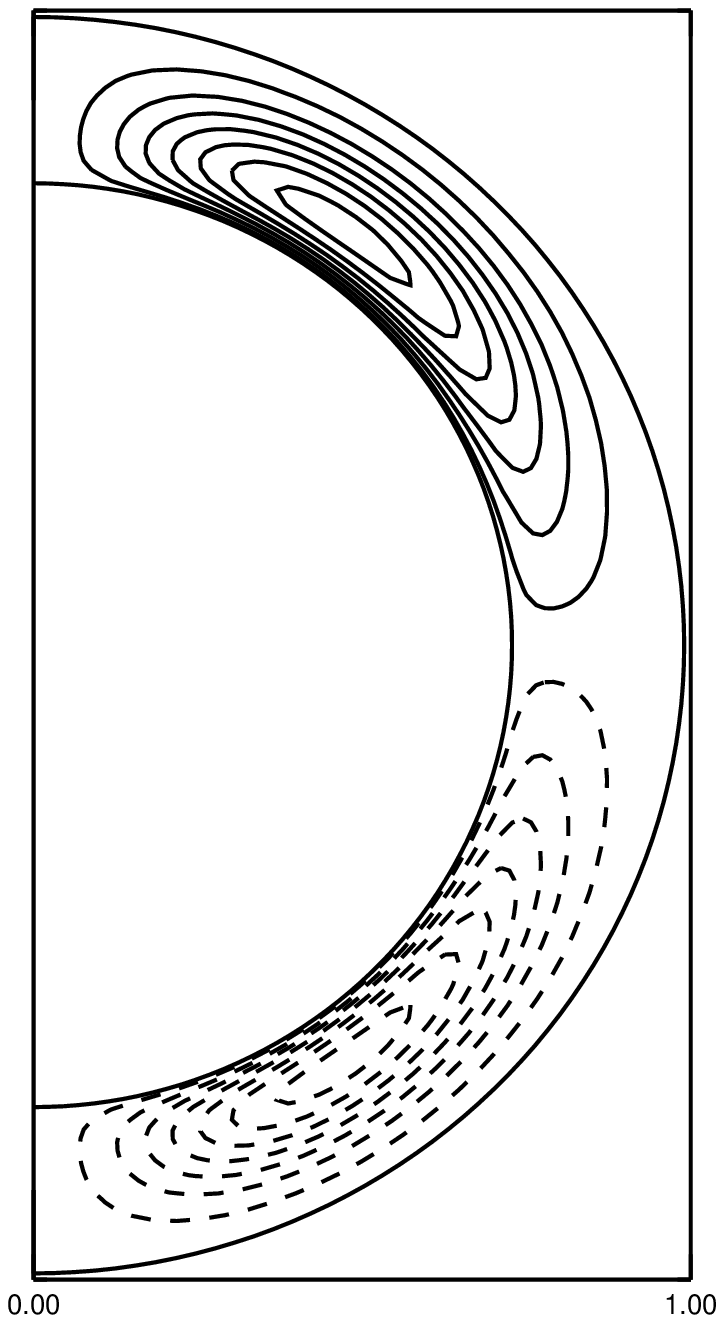}
 \includegraphics[width=3.0cm]{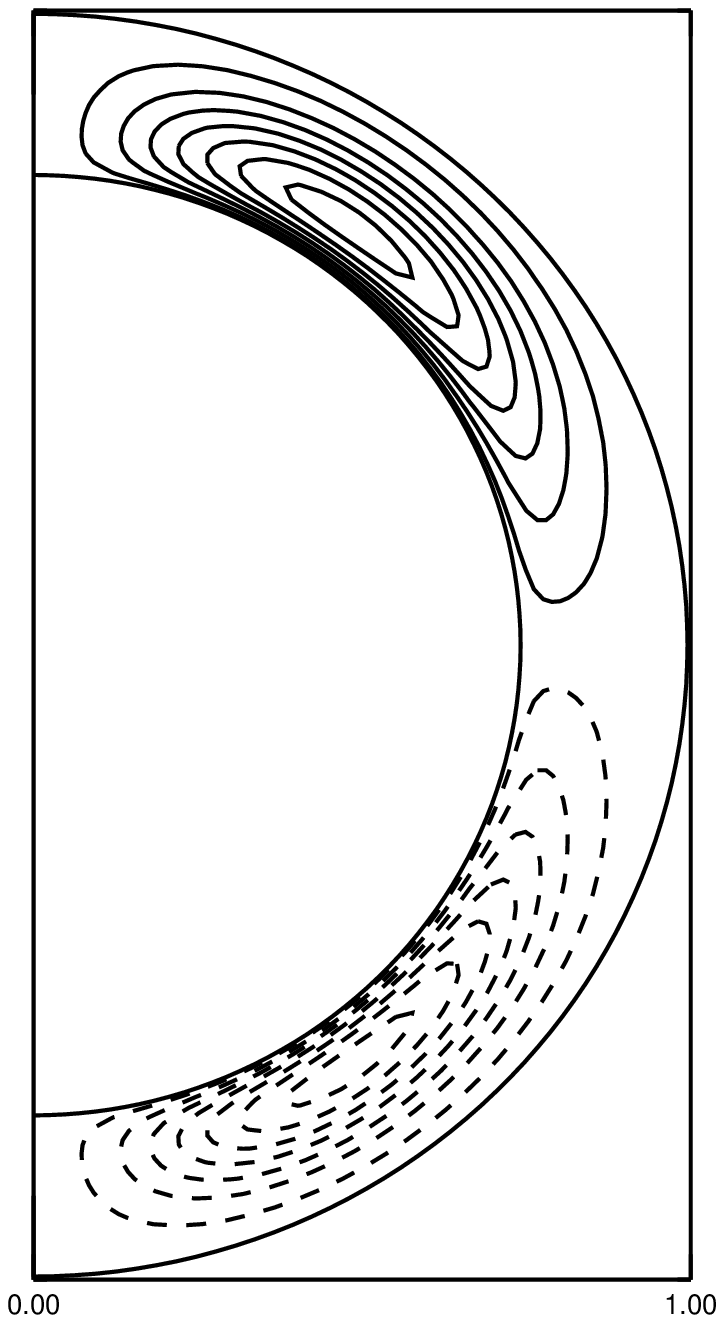} 
}
 \\ \mbox{
 \includegraphics[width=3.0cm]{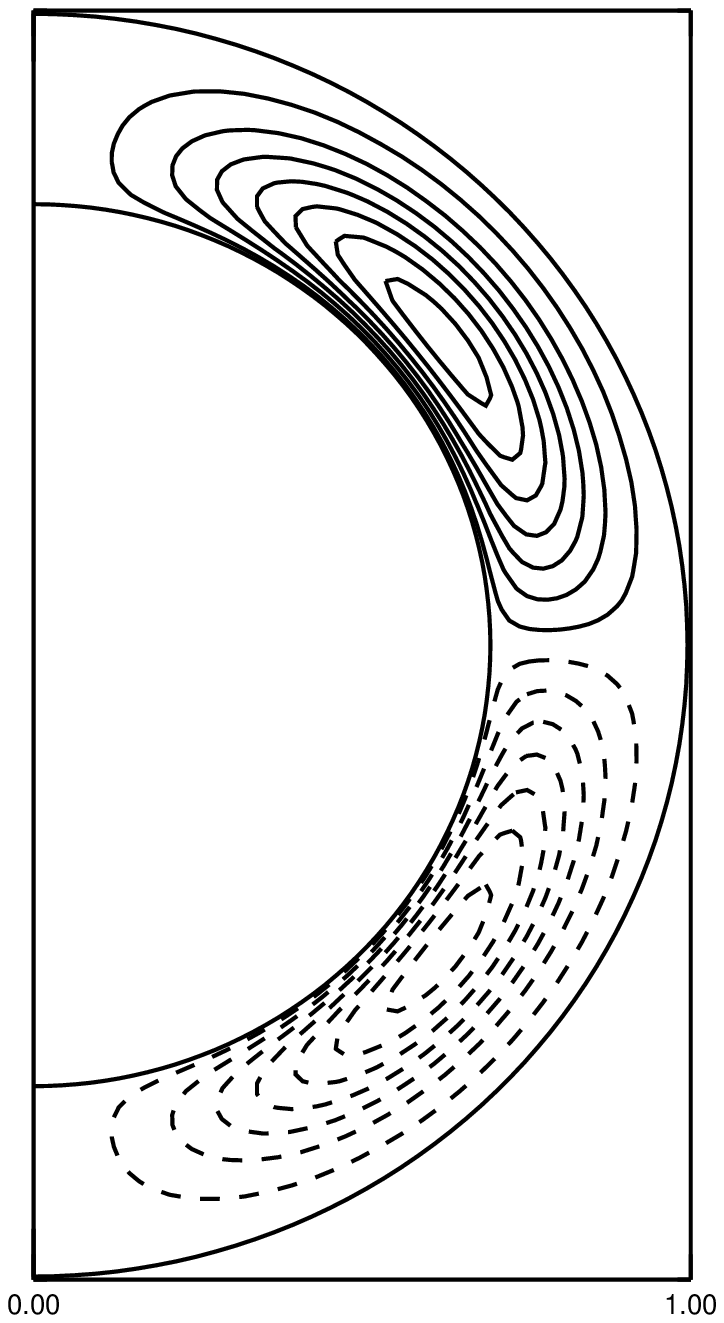}
 \includegraphics[width=3.0cm]{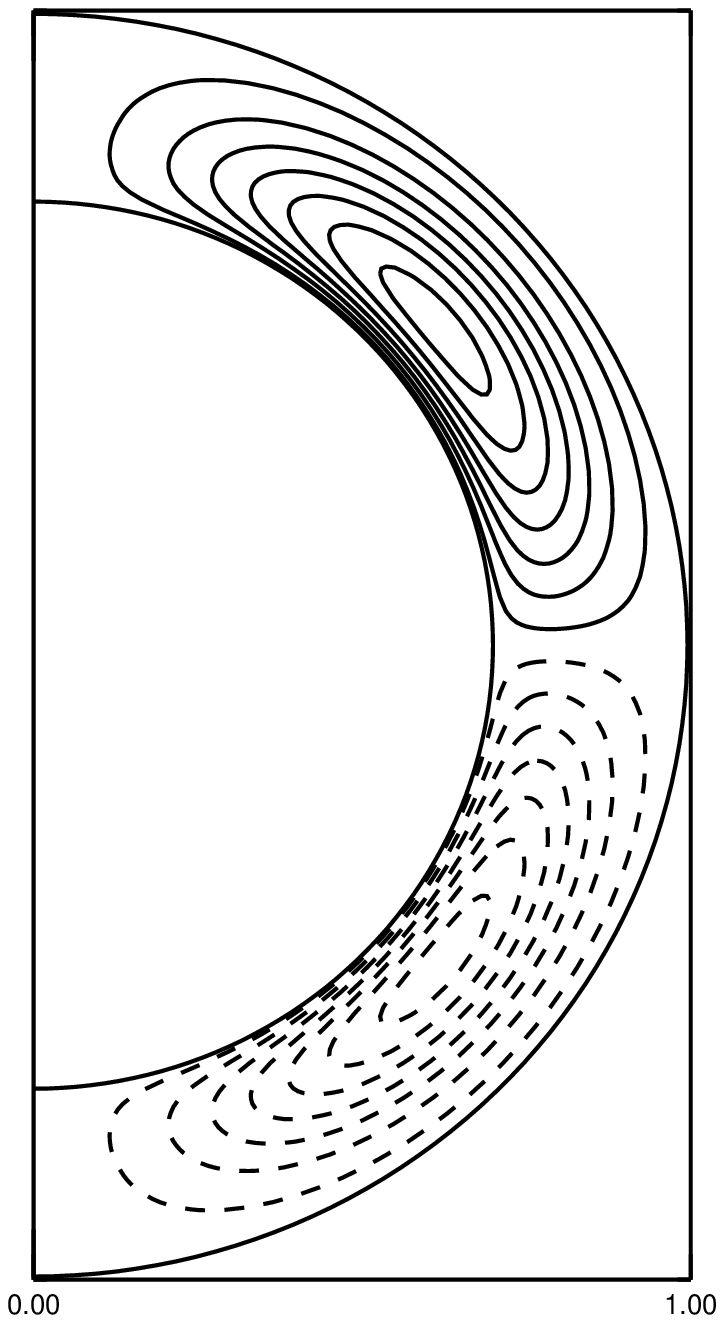} 
}
\caption{ \label{p33stream} The same as in Fig. \ref{p20stream} but for a rotation period 
$P_{\rm rot} = 33\,\rm d$. Here the meridional motion maximum speeds at the stellar surface are respectively: $65\,\rm m\,\rm s^{-1}$ for Model 1; $57\,\rm m\,\rm s^{-1}$ for Model 2; $17\,\rm m\,\rm s^{-1}$ for Model 3; $16\,\rm m\,\rm s^{-1}$ for Model 4.}
\end{figure}
%
\begin{figure}
\begin{center}
\includegraphics[width=8.5cm,height=8cm]{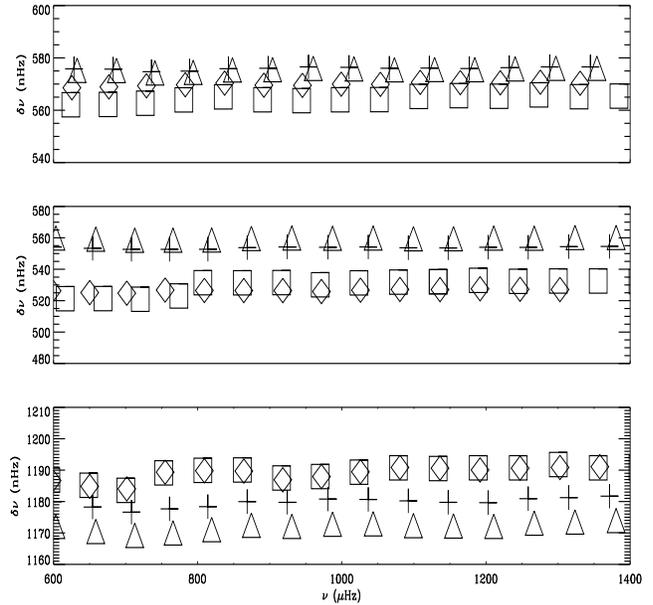}
\caption{The frequency rotational splittings of acoustic oscillation modes, $\delta\nu$, $l=1,\,m=1$ (top panel) 
$l=2,\,m=1$ (middle panel) and $l=2,\,m=2$ (bottom panel), 
as functions of the mode frequency, $\nu$,
for the four models of differential rotation with a rotation period $P_{\rm rot} = 20\,\rm d$. The symbols identify the models: diamonds: Model 1; squares: Model 2;
triangles: Model 3; crosses: Model 4.}
\label{split}
\end{center}
\end{figure}
The results of the present investigation are summarized in Figs. \ref{p20} -- \ref{split}. Figure \ref{p20} describes, for each of the four computed Procyon models, the behaviour of the normalized rotation as a function of fractional radius at different latitudes, $\phi = 0^\circ,\, 30^\circ,\, 60^\circ,\, 90^\circ$, for a rotation period $P_{\rm rot} = 20\,\rm d$. The normalization of angular velocity is done in terms of a weighting function based on the moment of inertia:
\begin{displaymath}
 \bar{\Omega} = \frac{\int \rho r^4 \sin^3 \theta \Omega { d}r { d}\theta}{\int \rho r^4 \sin^3 \theta { d}r { d}\theta}
\end{displaymath}
with the rotation period $P_{\rm rot} =  2\pi/\bar{\Omega}$.
Figure \ref{p20stream} describes, for the same models and rotation rate, the patterns of the meridional flow in the convection zone. 
Figures \ref{p33} and \ref{p33stream} show the same features of the Figs. \ref{p20} and \ref{p20stream}, but for a rotation period $P_{\rm rot} = 33\,\rm d$.  
Figure \ref{split} shows the rotational splittings, as deduced from the four differential rotation models of Procyon for $P_{\rm rot} = 20\,\rm d$, calculated for the acoustic oscillation modes with $l=1,\, m=1$ and $l=2,\, m=1,2$. 
 
In general, the overall characteristics of distribution of angular velocity in all the models are of solar type, with an equatorial acceleration and the surface rotation essentially maintained throughout the whole convection zone. 

The total horizontal shear at the surface, $\tilde\Omega = (\Omega_{\rm eq} -\Omega_{\rm pole})/\bar{\Omega}$, ranges from 0.1 to 0.45, depending on models and rotation period, that is of the same order of the solar one, $\tilde\Omega_\odot \simeq 0.25$. The surface shear appears to be stronger in the higher rotation case ($P_{\rm rot} = 20\,\rm d$). 
For Models 1 and 2, independently of their rotation rate, the rotation latitudinal shear is more pronounced at the surface than at the base of the convection zone, where there is a tendency to uniform rotation, as in the case of the Sun (Figs. \ref{p20} and \ref{p33}).
 
The isorotation surfaces, not presented here, are practically the same for all the models and the two rotation rates considered, and resemble the solar ones, nearly radial or slightly inclined with respect to the equatorial plane, certainly not of cylindrical shape, even in the lowest part of the convection zone. 
In most cases the meridional flow is in a single cell extending from the poles to the equator, with the surface motion directed toward the equator, except for the Models 1 and 2, at their higher rotation rate, where an additional small cell appears near the equator with the surface motion direct toward the pole (Fig. \ref{p20stream}). The maximum speed of meridional motions at the surface ranges from $16\,\rm m\,s^{-1}$ to $100\,\rm m\,s^{-1}$, depending on the models and rotation period (Figs. \ref{p20stream} and \ref{p33stream}). For comparison, the Sun's meridional flow at the surface is directed towards the poles with a speed of $\sim 10\,\rm m\,s^{-1}$ at intermediate latitudes, as deduced from Doppler shift observations (Hathaway 1996).

As far as the rotational splittings are concerned, the results reflect essentially the different dynamical behaviours of the Models 1 and 2, which have deeper convection zones and stronger horizontal rotational shears, with respect to those of Models 3 and 4, which have shallower convection zones and weaker horizontal rotational shears. For the modes with $m=1$ (top and middle panels of Fig. \ref{split}) there is a clear tendency of splittings related to the models with weaker rotational shear to approach the rigid rotation frequency of $579\,\rm nHz$, corresponding to $P_{\rm rot}=20\,\rm d$, while the splittings related to the models with stronger rotational shear are slightly smaller ($560-570\,\rm nHz$), in the case of $l=1$, and smaller ($520-530\,\rm nHz$), in the case of $l=2$, than the rigid rotation frequency.  Opposite is the case of modes with $l=2,\, m=2$ (bottom panel of Fig. \ref{split}) in which the splittings frequencies related to the models with weaker rotational shear appear to be smaller than those related to models with strong rotational shear. The origin of the differences in the splitting for models 1, 2 and 3, 4 comes mostly from the rotational profile, in particular from the latitudinal dependence of the shear. In fact, if the latitudinal dependence of the
rotational law is artificially suppressed by setting $i_{max}=0$ in Eq.(16), the differences in the splittings turn out to be much smaller. The fact that a stronger rotational shear lowers the average frequency of splittings for modes with $m=1$, and viceversa enhances the frequency for modes with $m=2$ depends mainly from the contributions of the angular function terms in kernels of Eq.(17), which are negative for $m=1$ and positive for $m=2$ in our models.
As a consequence, the sum in Eq.(17) is lowered in the first case and enhanced in the second case, this trend being more evident for models 1,2 where the latitudinal shear is stronger.

%
\section{Conclusions}
%
Procyon has already been observed extensively and will presumably be observed also in the future with a larger accuracy both from ground, by means of multi-site campaigns, or from space. Therefore more accurate determinations of the basic parameters of the star as well as a higher resolution spectrum of its acoustic pulsations are expected, such that can be useful to verify the results of the present investigation.

We have shown that a differential rotation of solar type can be maintained even in a very shallow convection zone and that its characteristics are enough sensitive to permit the discrimination among different stellar models; Models 1 and 2, which have a more extended convective envelope, show a larger latitudinal shear than the Models 3 and 4, which have a less extended convective envelope. For the models with the higher rotation rate, the meridional motion at the lower latitudes changes from poleward, for Models 1 and 2, to equatorward, for Models 3 and 4. The features of the surfaces of isorotation seem to be adequate to ensure a possible cyclic dynamo activity in this star, whose surface manifestations could depend both on differential rotation and meridional flow.

Precise measurements of the rotational splittings should permit in principle the discrimination between models with deeper convection zones and strong latitudinal shears (Models 1 and 2) with respect to those with shallower convection zones and  weak latitudinal shears (Models 3 and 4). 
On the basis of the present study, a resolution of $\sim 10\,\rm nHz$ in the oscillation spectrum seems to be necessary to test our models of Procyon. 
Although this resolution is barely reachable for individual oscillation frequency measurements in the near
future, even with the forthcoming space missions, since the mean value of the splitting for each mode is not strongly dependent on the frequency, the use of averages over several measured splittings could permit to reach the required accuracy.

%

\end{document}